\documentstyle[12pt,epsfig]{article}
\begin{document}
\sloppy
\title{Quark effects and
 isospin-violation in ${}^1S_0 \,\,NN$ scattering}
\author{Chung Wen Kao and Shin Nan Yang  \\
 Department of Physics,
National Taiwan University\\ Taipei, Taiwan 10617} \maketitle

\begin{abstract}
It has long been established that the isospin symmetry is slightly
broken in the nuclear interaction. On the quark level, most
studies on the violation of isospin symmetry focus on the mass
difference between $u$ and $d$ quarks, in addition to the (direct)
Coulomb interactions between quarks.
 However, it has been demonstrated that there is other source of isospin
violation on the quark level. Maltman, Stephenson, and
Goldman(MSG) showed that the interference effects of QED and QCD
gave significant contributions to the baryon isomultiplet mass
splittings and binding energy difference between ${}^3H$ and
${}^3He$. In this work, we present the results of
the effects of these new mechanisms on isospin violation in $NN$ ${}^1S_{0}$
scatterings with the new sets of
strength parameters obtained by
fitting to the mass splittings in the baryon isomultiplets listed
in the most recent PDG compilation. We also give the matrix elements of the
various potential operators which would be useful in the study of
these charge dependent effects in finite nuclei.
\end{abstract}

\newpage
\noindent
{\bf 1. Introduction}\\

 It is well established that isospin symmetry is slightly broken in the
nuclear interaction. However, isospin violation still remains as
one of the lesser understood aspects of the nuclear force. 
Both the charge independence breaking (CIB)
and charge symmetry breaking (CSB) effects have been extensively
 studied within the hadronic picture (for recent reviews,
see Ref. \cite{Miller90}). For example,  
$\pi^{\pm}$ and $\pi^0$ mass difference
in the one-pion-exchange potential (OPE) and the exchange of a $\gamma\pi$
pair \cite{chemtob75,Yang83,Friar96} break the charge independence, while
meson mixings like $\pi-\eta$ and $\rho-\omega$  induce charge symmetry
breaking interactions \cite{GHT,PW,HHMK,KWW,OPTW}. \\

In addition to the interactions induced by the exchange of heavier
mesons, the short-range isospin violating force could also arise from
the quark effects. Well-known sources of isospin violation on
quark level  include the $u-d$ quarks mass difference and the
electromagnetic interaction (one-photon-exchange) between quarks.
Their effects on the charge dependence of the $NN$ scattering
length were investigated within the constituent quark model in
\cite{Chemtob83,BFH85,WWW88} and found to be important but model
dependent. The $u-d$ mass difference effects on
Okamoto-Nolen-Schiffer anomaly \cite{WWL88} and isovector mass
shifts in nuclei \cite{Nakamura96} have also been studied and
found to be large. Several years ago, Maltman, Stephenson, and
Goldman (MSG) \cite{MSG90}  proposed a new source of isospin
violation on the quark level, namely, the interference effect of
QED and QCD, i.e., the penguin-like and box-like diagrams as
depicted in Fig. 1 and 2. They demonstrated that this mixed QED-QCD
effect gave significant contributions to the baryon isomultiplet
mass splittings \cite{MSG90} and binding energy difference between
$^3He$ and $^3H$ \cite{MGS91}. Another isospin violation
interaction on the quark level is associated with the quark-pion
coupling. Besides the conventional long range one-pion-exchange
potential (OPEP), the quark-pion coupling also generates short-range
repulsion and medium-range attraction from the exchange of quarks from
two different nucleons \cite{Liu93}. When the mass difference
between charged and neutral pions $\Delta  m_{\pi} = m_{\pi^{\pm}}
- m_{\pi^0} $ is taken into account, it induces a charge dependent
quark-exchange OPE (CDEOPE) interaction.\\

The difference between proton-proton and proton-neutron singlet
S-wave (${}^1S_{0}$) scattering lengths is a good testing ground
to study isospin violation
 because the ${}^1S_{0}$ two-nucleon system is nearly bound so that the
scattering length is large and very sensitive to the details of the
$NN$ potential. 
We have studied contributions of the QED-QCD mixing effect
 and CDEOPE interaction
to the charge dependence of the scattering length difference in
$^1S_{0}$ $NN$ scattering within a nonrelativistic constituent
quark model and the preliminary results were reported in Refs.
\cite{KY1,KY2}. In this paper, we  present the details of our
calculation and the results of a more extensive study on these
quark effects on the charge dependence of the scattering length
difference in $^1S_{0}$ $NN$ scattering. Furthermore, the new sets of
strength parameters of the QED-QCD mixing potential, obtained by
fitting to the mass splittings in the baryon isomultiplets listed
in the most recent PDG compilation \cite{PDG}, are used in the
current calculation. We also give the matrix elements of the
various potential operators which would be useful in the study of
these CD effects in finite nuclei. \\

  In Sec. 2, we discuss the quark mediated charge dependent interaction.
We focus on QED-QCD mixing effects and the charge dependent
potential associated with the pion exchange between quarks when the
$\pi^{\pm}-\pi{^0}$ mass difference is taken into account. In Sec. 3
the determination of parameters in the QED-QCD mixing effects is discussed
in details and the new sets of strength parameters are obtained with
the use of a new mass splitting value of  $\Sigma_c^{++} - \Sigma_c^{0}$.
 The resonanting group scattering equation used to estimate their effect
 on the $NN$ S-wave(${}^1S_{0}$) scattering length is derived in Sec. 4. The
results are presented and compared with the experimental values in Sec. 5
and the summary is also given there.\\

\noindent {\bf 2. Quark mediated charge dependent $NN$
interactions}\\

The aim of this paper is to investigate the various quark
effects on the isospin violation in
$^1S_{0}$ $NN$ scattering.
Quark-mediated isospin violation mechanisms considered here
 include $u-d$ quark mass difference,
quark-exchange Coulomb interaction, QED-QCD mixing effects and pion mass
difference in quark-exchange OPE interaction. Each effect will be
individually discussed below.\\

\noindent
2.1. Up-down quark mass difference effect\\

The mass difference $\delta m_q \equiv m_d - m_u$ between $u$ and $d$ quark
 induces isospin violation
of $NN$ interaction because it explicitly breaks the $SU(2)$ isospin
symmetry of QCD Lagrangian.
In the constituent quark model, quark mass difference not only breaks
 isospin symmetry through the kinetic energy of constituent
quarks, but also through one-gluon-exchange (OGE) interaction
between quarks. In $^1S_0$, only the Fermi contact hyperfine
interaction contributes to the isospin violation of $NN$ scattering:
\begin{equation}
V_{glu}=\sum_{i<j=1}^{6}-(\frac{\vec{\lambda}_{i}}{2}\cdot\frac{\vec{\lambda}_
{j}}{2})
(\frac{\vec{\sigma}_{i}}{2}\cdot\frac{\vec{\sigma}_{j}}{2})\frac{8\pi\alpha_{s}}
{3m_{i}m_{j}}\delta^{(3)}(r_{ij}),
\end{equation}
where $\alpha_{s}$ is the strong coupling constant.
 The $\Delta N$ splitting $m_{\Delta}-m_{N} =
 \frac{4 \alpha_{s} \alpha^{3}}{3 \sqrt{2 \pi}m_{0}^{2}}
 \simeq 260$ $MeV$ leads to $\alpha_s$=$1.624$, where $m_0=330 MeV$
 is the average mass of $u$ and $d$ quark. \\

\noindent
 2.2. Exchanged pion mass difference effect\\

 Another important interaction between the quarks arises from the
quark-pion coupling.  Quark-pion coupling is known to be very
important to ensure chiral symmetry in quark models. It has been
used to study various aspects of the $NN$ interaction, for
example, the conventional long range one-pion-exchange potential
(OPEP). In addition, it can also generate short-range repulsion
and medium-range attraction from the exchange of quarks from two
different nucleons \cite{Liu93}.
 When the mass difference between charged and neutral pions
 $\Delta m_{\pi}=m_{\pi^{\pm}} - m_{\pi^{0}}$ is taken into account, it gives rise
to the following charge dependent one-pion-exchange interaction
between quarks $i$ and $j$,
\begin{eqnarray}
V^{opep}_{qq}(\Delta m_{\pi}) &=& \frac 13 (\frac {f^2_{\pi
qq}}{4\pi}) \{ m_{c} ( \vec\tau_i \cdot \vec\tau_j) \{ (
\vec\sigma_i \cdot \vec\sigma_j) +
 [ 1+ \frac {3}{m_{c} r} + \frac {3}{m_{c}^2 r^2} ]
 S_{ij} \}
 \frac {e^{-m_{c} r}}{m_{c} r} \nonumber\\
 &-& \tau_{iz}\tau_{jz} \{ m_{c} \{ (\vec\sigma_i \cdot \vec\sigma_j) +
 [ 1+ \frac {3}{m_{c} r} + \frac {3}{m_{c}^2 r^2} ] S_{ij} \}
 \frac {e^{-m_{c} r}}{m_{c} r} \nonumber\\
 &-&  m_{0} \{ (\vec\sigma_i \cdot \vec\sigma_j) +
 [ 1+ \frac {3}{ m_{0} r} + \frac {3}{ m_{0}^2 r^2} ] S_{ij} \}
 \frac {e^{- m_{0} r}}{ m_{0} r}\}\},
\end{eqnarray}
where $m_{c}= m_{\pi^{\pm}}$, $m_{0} = m_{\pi^{0}}$ and $S_{ij} =
3 (\vec\sigma_i \cdot \hat{r}_{ij}) (\vec\sigma_i \cdot
\hat{r}_{ij})- \vec\sigma_i \cdot \vec\sigma_j$ and $r=|\vec
r_{ij}|$.\\

Identifying the folded OPEP with the familiar OPEP between
nucleons gives $f^2_{\pi qq} = \frac {9}{25} f^2_{\pi NN}
e^{-\frac {m_{\pi}^2}{3\alpha^2}}$ \cite{Chemtob83}, with
$f^2_{\pi NN}/4\pi = 0.079$.\\

\noindent
 2.3. Exchanged electromagnetic quark effect\\

 The electromagnetic
 interaction is an
 important mechanism of isospin violation.
 The one-photon-exchange diagram induces both Coulomb potential and
 hyperfine interaction:
\begin{equation}
 V_{\gamma}=V_{\gamma}^{c}+V_{\gamma}^{hypf},
\end{equation}
where
\begin{eqnarray}
 V_{\gamma}^{c}&=&\sum_{i<j=1}^{6}\frac{\alpha_{em}Q_{i}Q_{j}}{r_{ij}}, \\
V_{\gamma}^{hypf}&=&\sum_{i<j=1}^{6}-Q_{i}Q_{j}(\frac{\vec{\sigma}_{i}}
{2}\cdot\frac{\vec{\sigma}_{j}}{2})\frac{8\pi\alpha_{em}}{3m_{i}m_{j}}
\delta^{(3)}(r_{ij}).
\end{eqnarray}
\noindent Here $Q_{u}  = 2/3$, $Q_{d} = -1/3$ and
 $\alpha_{em} =  1/137$.
Note that direct Coulomb effect should be subtracted. However
Coulomb effect between two exchanged quarks has to be taken into
 account. \\

\noindent 2.4. QED-QCD mixing effect\\

The QED
 and QCD mixing effect, as suggested by MSG, refers to those of the
penguin-like and box-like diagrams as depicted in Figs. 1 and 2,
respectively. The penguin-like diagrams are the electromagnetic
vertex correction to the one-gluon exchange graphs. The box-like
diagrams are attributed to those of the gluonically dressed
versions of the basic one-photon exchange graphs.
 In general, both of them include graphs to all orders of $\alpha_s$ while
only the lowest-order graphs in $(\alpha\alpha_s)$ are shown in
Figs. 1 and 2. MSG based their argument on two chiral constraints
on the pion electromagnetic (EM) self-energy and demonstrated that
these additional terms should be present. The first chiral
constraint is that
 \cite{Gasser85} the
contribution from up-down quark mass difference, $\delta m_q$, to
the $\pi^{\pm} - \pi^0$ splitting is of order $O((\delta m_q)^2)$
and small (less than $0.17 \, MeV$) for physical light-quark
masses. The $\pi^{\pm} - \pi^0$ splitting is, therefore, due
essentially entirely to pion EM self-energy. 
The second is, in the chiral
limit \cite{Dashen69} $(\delta m_{\pi^0}^2)_{em} = 0$, and the
correction to the chiral expansion of this constraint is $(\delta
m_{\pi^0})_{em} <0.5\,\,MeV$
 \cite{Maltman90}. Hence the  entire $\pi^{\pm} - \pi^0$ splitting
should come from $\pi^{\pm}$  EM self-energy. However, this cannot
be the case within the constituent quark model if the
one-photon-exchange potential is taken to be the only mechanism of
electromagnetic interaction between quarks. The operator which
corresponds to one-photon-exchange graph has the form of $Q_1 Q_2
F$, where $F$ includes all the other color, spin and spatial
dependences. The expectation values of the charge dependent operator $Q_1
Q_2 F$ in $\pi^{\pm}$ and $\pi^0$, are respectively, $2x/9$ and
$-5x/9$, where $x =\,
\langle F \rangle_{\pi}$. 
The chiral constraint of $(\delta m^2_{\pi^0})_{em} = 0$ would lead
to $(\delta m_{\pi^{\pm}})_{em} = 0$ and $(m_{\pi^{\pm}} -
m_{\pi^0}) < 0.17 \,\,MeV$, in contradiction with the experiment.
On the other hand, if we insist in $ (m_{\pi^{\pm}} -
m_{\pi^0})_{em} \simeq (m_{\pi^{\pm}} - m_{\pi^0})_{exp},$ then
$(\delta m_{\pi^0})_{em} \simeq -\frac {5}{7} (m_{\pi^{\pm}} -
m_{\pi^0})_{exp} \neq 0$, and is inconsistent with the chiral
constraint. To resolve this problem, MSG asserted that it was
necessary to include the penguin-like and box-like diagrams.
The explicit forms of their effective potentials and associated 
coefficients are discussed in Sec. 3.\\

\noindent
{\bf 3. The Hamiltonian}\\

Our calculation is performed in a non-relativistic
constituent quark model with
 a six-quark Hamiltonian,
 constructed as  a sum of one-body
kinetic energy, mass and
 two-body potential energy \cite{Chemtob83}.
It is given as:\\

$$H={\cal K}+M+V_{conf}+V_{glu}+
V_{qq}^{opep}+V_{\gamma}+V_{box}+V_{penguin}.$$
Since the constituent quark is heavy, we use the non-relativistic form for the
free energy of the quarks and write

$${\cal K}+M=\sum_{i=1}^{6}(\frac{p_{i}^{2}}{2m_{i}})+6m_{i}. $$
Here, $m_{i}$ is the mass of i-th quark and we use
 $m_{u}= 330 MeV $. The confining potential is taken to be

$$V_{conf}=\sum_{i<j=1}^{6}-(\frac{\vec{\lambda}_{i}}{2}\cdot
\frac{\vec{\lambda}_{j}}{2})(\frac{3}{4}Kr_{ij}^{2}+e_{0}),$$
For simplicity, the anharmonic component is not included in $V_{conf}$.
The oscillator length $\alpha
= (3m_{0}K)^{1/4} = 320 MeV = 0.617 fm^{-1}$, and the constant
 $e_{0} =  -385 MeV$ are all fixed by the hadron spectra.\\

The other terms $V_{glu}$, $V_{qq}^{opep}$, $V_{\gamma}$,
 are the charge dependent two-body
potential terms. Their forms and parameters have already been
given in Sec 2. Following we only discuss the terms associatedwith
QCD-QED interference effects.
Four effective potential operators are introduced by MSG to
account for the effects of those penguin-like and box-like
diagrams:
\begin{equation}
V_{box}=V_{box}^{c}+V_{box}^{hypf},
\end{equation}
where
\begin{eqnarray}
V_{box}^{c}&=&\sum_{i<j=1}^{6}A(\frac{\vec{\lambda}_{i}}{2}\cdot\frac{\vec
{\lambda}_{j}}{2})\frac{\alpha_{em}Q_{i}Q_{j}}{r_{ij}},\\
V_{box}^{hypf}&=&\sum_{i<j=1}^{6}B(\frac{\vec{\lambda}_{i}}{2}\cdot\frac{\vec{
\lambda}_{j}}{2})Q_{i}Q_{j}(\frac{\vec{\sigma}_{i}}{2}\cdot\frac{\vec{\sigma}_{
j }}{2})\frac{8\pi\alpha_{em}}{3m_{i}m_{j}}\delta^{(3)}(r_{ij}).
\end{eqnarray}
\noindent And
\begin{eqnarray}
V_{penguin}&=&V_{penguin}^{c}+V_{penguin}^{hypf},
\end{eqnarray}
where
\begin{eqnarray}
 V_{penguin}^{c}&=&\sum_{i<j=1}^{6}C(\frac{\vec{\lambda_{i}}}{2}\cdot\frac{\vec{
\lambda_{j}}}{2})\alpha_{em}\frac{(Q_{i}^{2}+Q_{j}^{2})}{r_{ij}},
\\
 V_{penguin}^{hypf}&=&\sum_{i<j=1}^6D(\frac{\vec{\lambda}_i}{2}\cdot\frac{\vec
 {\lambda}_j}{2})
(Q_{i}^{2}+Q_{j}^{2})(\frac{\vec{\sigma}_{i}}{2}\cdot\frac{\vec{\sigma}_{j}
} {2})\frac{8\pi\alpha_{em}}{3m_{i}m_{j}}\delta^{(3)}(r_{ij}).
\end{eqnarray}
\noindent
 $\vec{\lambda}_i$ ({\it i}=1, 2...8)
 are the color SU(3) group generators with standard normalization,
 $Tr(\lambda_i \lambda_j)=2 \delta_{ij}$. $A$, $B$, $C$, $D$ are the strength
parameters determined \cite{MSG90} from fitting to the
isomultiplet mass splittings in the baryon octet and decuplet, and
$\Sigma^{++}_c - \Sigma^0_c$ with the assumption that the
splittings are induced by the $u-d$ quark mass difference,
one-photon-exchange potential and the QED-QCD mixing effect
discussed above. \\

The mass splittings in the baryon octet and decuplet depend only
on $\delta m_q$, $\alpha$, $\beta$, and $\gamma$, but not on
$\delta$, where $\alpha = A\epsilon - C\epsilon$, $\beta = B\mu-
D\mu$, $\gamma = 4C\epsilon + D\mu$, $\delta =D\mu/4$
 ($\epsilon = \langle \frac{\alpha_{em}}{r_{ij}}
 \rangle_{nucleon}=1.86 MeV$,
 $\mu=\langle\frac{8\pi\alpha_{em}}{3m_{i}m_{j}}\delta^{(3)}(r_{ij})
 \rangle_{nucleon}=1.17 MeV$).
To determine $\delta$, it is necessary to go beyond SU$(3)_f$
baryons, i.e., to systems with different symmetry properties and
$\Sigma^{++}_c - \Sigma^0_c$ was chosen in Ref. \cite{MSG90} for
this purpose. The QED-QCD interference
effects on the charge dependence in $NN\,\, ^1S_0$ scattering
length reported in \cite{KY1,KY2} were obtained with $A$, $B$,
$C$, $D$ determined in \cite{MSG90} with 
the value of $\Sigma^{++}_c - \Sigma^0_c = 0.2\pm0.5\,
MeV$.
The most recent PDG \cite{PDG} compilation now gives
$\Sigma^{++}_c - \Sigma^0_c = 0.66 \pm 0.28 \, MeV$. We have
carried out a refitting procedure with this latest value of
$\Sigma^{++}_c - \Sigma^0_c$. 
The results are reported in Sec 5.\\

\noindent
{\bf 4. The resonanting group method}\\

 The variational scattering equation for the $NN$ system is derived by the
 resonanting group method (RGM). We briefly
 sketch the method here and refer the readers to Ref. \cite{Chemtob83} for
 details.
It is based on the Hill-Griffin-Wheeler variational principle, namely,
the minimization of the functional $\langle \Psi|H-E|\Psi\rangle$,
\begin{equation}
\langle\xi|\Psi\rangle=\Psi(\xi)=\sum_\alpha\int d\vec{R} g_\alpha(\vec{R}) \Phi_\alpha(\xi;\vec{R}),
\end{equation}
where the $6q$ state function $\Psi$ is a functional of the trial weight
function $g_\alpha(\vec{R})$ which depends on the collective variable
$\vec{R}$, and $\Phi_\alpha(\xi;\vec{R})$ denotes the $6q$ wave function in
configuration of two clusters A and B. $\alpha$ is a label for the quantum
numbers of the clusters and $\xi$ represents the internal coordinate sets. In
this study, we will restrict our consideration to the case where both
clusters A and B are in the nucleon ground state and write
\begin{equation}
  \Phi(\xi;\vec{R})={\cal A}
 [\phi_A \phi_B
 \delta^{(3)}(\vec{X}-\vec{R})],
\end{equation}
where $\phi$ is the ground state wave function of the $3q$ cluster
and $\cal A$ is the antisymmetrization operator, i.e., ${\cal A} =
\sqrt{\frac {1}{10}} (1-\sum_{i\in A, j\in B}P_{ij}),$ with
$P_{ij}$ the permutation operator for particle labels $i$ and $j$.
We treat all the interaction terms internal to the clusters,
except the harmonic potential in $V_{conf}$, perturbatively up to
the first order. This allows us to work with the cluster wave
functions in the form
\begin{eqnarray}
\phi_A& = &\phi_A^0(\vec\rho,\vec \lambda) \phi^c \phi^{\sigma\tau},\nonumber\\
\phi_A^0(\vec\rho,\vec \lambda)& = &\frac { \alpha^{3/2}_\rho
\alpha^{3/2}_\lambda}{\pi^{3/2}}
e^{-\frac 12(\alpha^{2}_\rho\rho^2+ \alpha^{2}_\lambda \lambda^2)},
\label{clusterwf}
\end{eqnarray}
where $\vec \rho$ and $\vec \lambda$ are the conventional Jacobi
coordinates in the $3q$ cluster A and $\alpha_\rho = (3mK)^{1/4}$,
$\alpha_\lambda = (3m_\lambda K)^{1/4}$, $m_\lambda =
3mm'/(2m+m')$, with $m$ the mass of the like quark and $m'$ the
mass of the unlike quark. $\phi_A^0$, the spatial part of
$\phi_A$, is the ground state harmonic-oscillator wave function,
 $\phi^c$ the $3q$ color singlet sate function and $\phi^{\sigma\tau}$ the
 flavor-spin state function of the nucleon. For simplicity, we will use the
 charge-symmetric limit form of Eq. (\ref{clusterwf}) by setting
$\alpha_\rho = \alpha_\lambda = (3m_0 K)^{1/4}.$ \\

The variational principle then gives
\begin{equation}
\int d\vec R' [H(\vec R,\vec R') - E A(\vec R,\vec R')] g(\vec R') =0, \label{RGM}
\end{equation}
where
\begin{eqnarray}
H(\vec{R},\vec{R'})&=&\langle\Phi(\vec{R})|H|\Phi(\vec{R'})\rangle\nonumber \\
&=&\langle\Phi^{0}(\vec{R})|H|\Phi^{0}(\vec{R'})\rangle-9\langle\Phi^{0}(\vec{R})
|HP_{lm}|\Phi^{0}(\vec{R'})\rangle, \label{Hkernel}
\end{eqnarray}
and
\begin{eqnarray}
A(\vec{R},\vec{R'})&=&\langle\Phi(\vec{R})|\Phi(\vec{R'})\rangle\nonumber \\
&=&\langle\Phi^{0}(\vec{R})|\Phi^{0}(\vec{R'})\rangle-9\langle\Phi^{0}(\vec{R})
|P_{lm}|\Phi^{0}(\vec{R'})\rangle,\label{Akernel}
\end{eqnarray}
with $\Phi^{0}(\xi;\vec{R}) =
 \phi_0(\vec{\rho}_A,\vec{\lambda}_B)
\phi_0(\vec{\rho}_B,\vec{\lambda}_B)
\delta^{(3)}(\vec{X}-\vec{R}).$ The subscripts $l, m$ of $P_{lm}$
in Eqs. (\ref{Hkernel}) and (\ref{Akernel})
 are understood to belong to  different clusters.
Both  kernels $H(\vec{R},\vec{R'})$ and $A(\vec{R},\vec{R'})$ are divided
into  a direct
(without $P_{lm}$) and  an exchange (with $P_{lm}$) part.  If we  define

\begin{equation}
\langle\Phi^0 (\vec{R})|HP_{lm}|\Phi^0 (\vec{R'})\rangle\equiv H_E(\vec{R},\vec{R'}),
\end{equation}

\begin{equation}
\langle\Phi^0 (\vec{R})|P_{lm}|\Phi^0 (\vec{R'})\rangle \equiv K(\vec{R},\vec{R'}),
\end {equation}
then we  can express Eq. (\ref{RGM})
in the form of a Schroedinger equation
 with a non-local potential, after the removal of the CM motion of the
 two clusters,

\begin{eqnarray}
 [-\frac{\bigtriangledown_{\vec{R}}} {2\mu_{AB}^{(0)}}-(E-\bar{M}_A-\bar{M}_B)
\nonumber+V^D_{opep}(\vec{R})+V^D_{\sigma}(\vec{R})
+V_{\gamma}^D({\vec{R}})] g(\vec{R})
 + \int S(\vec {R},\vec{R'})
g(\vec{R})d\vec{R'}=0, \label{RGM1}
\end{eqnarray}
where
\begin{eqnarray}
 S(\vec {R},\vec{R'}) =  -9H_E(\vec{R},\vec{R'})+9 E K(\vec{R},\vec{R'}).
\end{eqnarray}
$\bar{M}_{A}$ is the nucleon mass predicted in this model,
and $\mu^{(0)}_{AB}$ is the reduced mass
of $NN$ system,

\begin{equation}
\mu^{(0)}_{AB}=\frac{M^0_{A}M^0_{B}}{M^0_{A}+M^0_{B}}\simeq\frac{1}{4}(M^0_{A}+M^0_{B}),
\end{equation}
with $M^0_p=2m_u+m_d$, $M^0_n=2m_d+m_u$. We have included in the direct term of Eq.
(\ref{RGM1}) a contribution $V^D_{\sigma}$ which  arises from $\sigma-$quark coupling.
This is to take into account the observed medium-range attraction in the $NN$
interaction which has been associated with the exchange of low-mass weakly correlated
$2\pi$ states. The expressions for all the direct terms can be found in \cite{Chemtob83}. \\

 The next main task is to evaluate the exchange kernels of Eqs.
  (\ref{Hkernel}) and (\ref{Akernel}).  Again, the expression for
  $K(\vec R,\vec R')$ and the one-body part of $H_E(\vec R,\vec R')$
  have been presented in \cite{Chemtob83}. For the two-body part  of $H_E(\vec R,\vec
  R')$, we have here the exchange Coulomb (EC), QED-QCD mixing effects and CDEOPE interaction which were not considered in
Ref. \cite{Chemtob83}. As in \cite{Chemtob83}, we split the two-body exchange matrix
elements into four components,

\begin{eqnarray}
 \sum_{l<m} \Gamma_{lm}P_{ij} &\equiv &\Gamma_{A}+\Gamma_{B}+\Gamma_{C}+
\Gamma_{C'}\nonumber\\
 &=&[\Gamma_{ij}P_{ij}]+[4(\Gamma_{ii'}+\Gamma_{i'j})P_{ij}]+
[4\Gamma_{i'j'}P_{ij}]
+[(\Gamma_{i'i''}+\Gamma_{j'j''})],
\end{eqnarray}
where the four components, enclosed within brackets, correspond to the four distinct
cases for the relation between indices $l, m$ of the interacting pair and $i, j$ of the
transposed pair: like indices (A), partially like (B), unlike intercluster (C) and the
unlike intracluster (C'). Since the $6q$ wave function features a simple factorization
in color, space and isospin-spin, it allows us to discuss independently the overlap
matrix element associated with each of these variables. Then the only new operators which
are not contained in $V_{conf}$ and $V_{glu}$ and not considered before
in \cite{Chemtob83} are  the spatial
operators $1/r_{ij}$ and $e^{-\mu r_{ij}}/r_{ij}$.  Only the  exchange matrix elements
of $e^{-\mu r_{ij}}/r_{ij}$ are presented in Table 1 since  $1/r_{ij}$ is a special
case of $e^{-\mu r_{ij}}/r_{ij}$ by letting $\mu =0$. We can now write

\begin{eqnarray}
S(\vec{R},\vec{R'}) &= &9[(E-M_{A}^{0}-M_{B}^{0})K(\vec{R},\vec{R'}) -{\cal
K}^{E}(\vec{R},\vec{R'})-V_{conf}^{E}(\vec{R},\vec{R'})-
 V_{glu}^{E}(\vec{R},\vec{R'}) \nonumber\\
 &&-V_{\gamma}^{E}(\vec{R},\vec{R'})-V_{qq, OPEP}^{E}(\vec{R},\vec{R'})-V_{box}^{E}(\vec{R},\vec{R'})
-V_{penguin}^{E}(\vec{R},\vec{R'})].
\end{eqnarray}

In the partial wave expansion, Eq. (\ref{RGM}) becomes
\begin{eqnarray*}
&&\lefteqn{[\frac{1}{2\mu_{AB}^{(0)}}(-\frac{d^2} {dR^2}+
\frac{l(l+1)}{R^2})+V_{\pi}^D (R)+V_{\sigma}^D (R)} \\ &&
-(E-\bar{M}_A- \bar{M}_B)] g_l(R) +\int_0^{\infty}dR
'RR'S_{l}(R,R') g_{l}(R')=0
\end{eqnarray*}
The above integro-differential equation is solved iteratively.
At large distance, $g_{l}(\vec{R})$
should approach
\begin{eqnarray}
 g_{l}(\vec{R})&\stackrel{R \rightarrow \infty}{\longrightarrow}&(kR)
(cos\delta _l(k)j_{l}(kR)+sin\delta_l(k)n_{l}(kR))  \nonumber \\
&\sim &sin(kR-\frac{1}{2}{\it l}\pi+\delta_{{\it l}}(k)),
\end{eqnarray}
where $k^2=2\mu^{(0)}_{AB}(E-{\bar M}_A-{\bar M}_B)$ defines the
relative c.m momentum. The effective-range parameters are
calculated from: $$k\cdot \cot\delta_{0}(k)=-\frac{1}{\it a}+
\frac{1}{2} r_0k^2+{\cal O}(k^4),$$
 where ${\it a}$ is the scattering
 length and $r_0$   the effective range. To  obtain
the values of $\delta {\it a}$, we use the relation $\delta {\it
a}=-{\it a}_{exp}^{2}\delta (1/{\it a})$  with ${\it a}_{exp}={\it
a}(pp)_{exp}\simeq $$-17.0 fm$.\\

\noindent
{\bf 5. Results and Discussions}\\

 The results of $\delta
a_{CSB}=a_{pp}-a_{nn}$ and $\delta
a_{CIB}=\frac{1}{2}(a_{pp}+a_{nn})-a_{np}$
are summarized in Table 2--6.
Here we have
carried out a refitting procedure with this latest value of
$\Sigma^{++}_c - \Sigma^0_c= 0.66 \pm 0.28 \, MeV$ given by
he most recent PDG \cite{PDG} compilation. 
Comparing to the previous results
\cite{KY1, KY2} where $\Sigma^{++}_c - \Sigma^0_c= 0.2 \pm 0.5 \, MeV$
was used, 
it is found that
$\delta a_{CSB}$ decreases about $0.27 fm$ and
$\delta a_{CIB}$ decreases about $0.15 fm$.
It indicates that QED-QCD interference effect exhibts only mild dependence
on the value of $\Sigma^{++}_c - \Sigma^0_c$. 
\\

The range of parameters corresponds to
$-4.40 \,\,MeV >\Delta \Sigma^* (= \Sigma^{*+} - \Sigma^{*-})
> -5.8 \,\,MeV$ and $5.51 \,\,MeV> \delta m_{q}>\,\,-2.97 MeV$.
On the quark level, the values of total $\delta a_{CSB}$, ranged
from $1.35 fm$ to $1.56 fm$, are relatively insensitive to the
choice of parameter sets.
On the other hand, the values of $\delta a_{CIB}$ are
much more sensitive to the choice of parameters. 
To understand this difference one needs to study each isospin 
violation mechanism in more details. 
\\

 The dominant CSB effects on the quark level are
 QED-QCD interference and quark mass difference effects.
Although individually each effect is very sensitive to the chosen parameters,
(Note that quark mass difference effect is ranged from $-1.83 fm$
to $3.24 fm$ with respect to different parameter sets;
 the penguin-like diagrams contribution is ranged from
$3.91 fm$ to $-1.32 fm$ and the box-like diagrams contribution
is ranged from $-1.14 fm$ to $-0.74 fm$)
 their sums still keep approximately constant.\\

On the other hand, similar cancellations do not occur in the CIB case.
Since $V(\delta m_{q})$ and penguin-like diagrams are charge
independent, the sensitivity of $\delta a_{CIB}$
with repsect to the chosen parameters
 is completely due to the fact that the box-like 
diagrams contribute to $\delta
a_{CIB}$ ranging from $-2.82 fm$ to $-1.82 fm$. Furthermore,
the CIB mechanisms on the quark level which are independent of
chosen parameter sets include the quark-exchange Coulomb
interaction and quark-exchange OPE interaction. The quark-exchange
Coulomb (EC) effect is about $0.38 fm$, the quark-exchange OPE
effect (CDEOPE) is only $0.08 fm$. Their sum is only about $0.46 fm$,
much less than the box-like diagrams contributions. It explains why
our values of $\delta a_{CIB}$ are more sensitive to the parameters.
\\

Furthermore, let us compare our results with the experimental data. 
The ${}^1S_0$ proton-proton scattering length is very accurately
measured, but the subtraction of the direct EM interaction is
model dependent \cite{Sauer77}. A commonly quoted value is $a_{pp}=-17.0\pm 0.2
fm$. The corresponding value for neutron-neutron scattering as
determined from $\pi^{-}d\rightarrow \gamma nn$ is
$a_{nn}=-18.5\pm 0.4 fm$ \cite{Schori87}. Our values of quark contributions to the
CSB range from $1.35 fm$ to $1.56 fm$, are very close the
experiemtnal value.
However, Coon and Niskanen \cite{Coon96} found
that from CSB vertex corrections and mass differences of the
intermediate baryons, the two-pion exchange would
contribute to $\delta a_{CSB}$. This effect would be as large as $1.37 fm$.
The sum of the meson and quark contributions would then be larger than the
experiemtnal value of $1.5 fm$, but is still within the
uncertainty incurred in the subtraction of direct EM interaction
in $pp$ scattering \cite{Sauer77}.\\

The ${}^1S_0$ neutron-proton scattering length is $-23.78 \pm
0.001 fm$, therefore $a(pp-np)=\delta a_{CIB}
+\frac{1}{2}\delta a_{CSB} \simeq 6.78 fm$ \cite{Dumbrajs}. 
It was claimed
in Ref. \cite{Ericson83} this result can be explained by various
meson-mediate mechanisms. The dominant contribution is from
the pion mass difference
effect in OPEP, its size was $3.6 \pm 0.15 fm$.
Another effect is due to the $\gamma \pi$ exchange which was 
$1.1 \pm 0.4 fm$. Such
an optimstic picture can be called into questions with a closer
look. First, the pion mass
difference effect in OPEP only contributes to $a(pp-np)$ $2.64 \pm 0.16 fm$
according to Refs. \cite{Yang83, Coon82} and confirmed by
the calculation of the Ref. \cite{Cheung86}.
For the $\gamma\pi$- exchange effect mediated by the seagull
vertex, Ref. \cite{chemtob75,Yang83} 
obtained only a contribution
of $0.67 \pm 0.02 fm$ to $a(pp-np)$ 
through the solution of
Schr\"odinger equation with RSC potential. A recent calulation of
the same effect even gave a result of $-0.35$ to $-0.53 fm$ with
opposite sign \cite{Friar96}. Taking these two together would
amount to a reductionof about $1.40$ to $2.60 fm$ from the total
estimate in Ref. \cite{Ericson83}. 
The quark contribution to $a(pp-np)$
in our calculation is ranged from
$-1.68$ to $-0.57 fm$.
It clearly makes the agreement between experiments and theories worse
since it is opposite in sign to the experimental value. If one
uses the the estimate of Refs. \cite{Yang83,Cheung86,Coon82} for
the pion mass difference effect in OPEP, Ref \cite{Friar96} for
$\gamma\pi$- exchange and Ref. \cite{Ericson83} for all other
meson-mediated effects respectively, and adds our values of quark
 contributions, then $a(pp-np)$ ranges from $1.31$ to $2.42
 fm$, well below the experimental value.\\

Although we have calculated $\delta a_{CIB}$(CDEOPE) and found its
value only about $0.08fm$, its contribution may be underesteemed.
Stancu {\it et. al.} \cite{SPG} pointed out that the
single channel Resonating Group Method maybe not adequate for the operators
with this factor
:$(\vec{\tau}_{i}\cdot\vec{\tau}_{j})(\vec{\sigma}_{i}\cdot
\vec{\sigma}_{j})$.
 Because the dominant configuration is no longer
$|s^{6}[6]_{o}[33]_{FS}\rangle$ but
$|s^{4}p^{2}[42]_{o}[51]_{FS}\rangle$, the reason is
because of its specific
flavor-spin symmetry. It requests further studies to esteem
more accurately on the size of this effect. \\

Furthermore, in the previous fit, the
 $\pi$-quark coupling inside the hadron has been completely
neglected since we truncated the $\pi$-quark coupling in the long and
medium range. To be consistent, quark-exchange OPE effect should
be also included in the fit of isomultiples of baryon spectrum if
$\delta a_{CIB}$(CDOPE) is taken into account. Such a refit is expected
to provide completely different values of all parameters here, including $\alpha_{s}$,
$\delta m_{q}$ and A,B, C, and D. However the recent calculation of Shih 
\cite{Shih} showed the $\pi$-quark couplings inside hadron are
small effects and refitted $\delta m_{q}$ increases only by $0.3 MeV$, therefore
we expect the new sets of parameters generated from this refit will
not dramatically differs with ours. \\

In conclusion,
the new sets of
strength parameters of the QED-QCD mixing potential, obtained by
fitting to the mass splittings in the baryon isomultiplets listed
in the most recent PDG compilation \cite{PDG}, is used in the
current calculation.
The effects of interference of QED and QCD on the
isospin-violation of ${}^1S_0 \,\,NN$ scattering are significant.
and very sensitive to the chosen parameters.  
The effects
of QED-QCD interference should be also significant
in charge-dependence in nuclear many-body systems such as $u$ and $d$ quark
mass difference effect does \cite{Nakamura96}, therefore
, we give the matrix elements of the
various potential operators which would be useful in the study of
these charge dependent effects in finite nuclei.
The charge dependent $NN$ potential
generated from several mechanisms on the quark level
explains the CSB effect of ${}^1S_0 \,\,NN$ scattering very well,
without including the questionable $\rho-\omega$ mixing effect. But
the CIB effect on the quark level carries the opposite sign
with respect to the experimental value and the agreement between
theory and experiment is still unavaible, so that 
the charge independence breaking in the NN interaction 
remains as an open issue in the quark models.
\\

{\bf Acknowledgment:}
This work was supported by the National Science Council/ROC under grant NCS
88-2112-M002-015.

\newpage
\begin{table}
\begin{center}
\begin{tabular}[thb]{|c|c|}\hline
\hline Type&Function \\ \hline
 &\\
 $A$ & $\frac{2}{3}\frac{\exp
(-\frac{3}{2}\mu|{\bf R-R'}|)}{|{\bf R-R'|}}$  \\
 &\\ \hline &\\
 $B$ &
 $\frac{\exp(\frac{5\mu^2}{16\alpha^2})}{\frac{3}{4} |{\bf
3R-R'}|}\{\exp[{-\frac{3 \mu}{8}|{\bf 3R-R'}|]
erf(\sqrt{\frac{4}{5}}\alpha[\frac{3}{8}|{\bf
3R-R'}|-\frac{5\mu}{8\alpha^2}]) +\exp(\frac{3 \mu}{8}|\bf
3R-R'}|)$ \\ $ $ &$ erf(\sqrt{\frac{4}{5}}\alpha[\frac{3}{8}|{\bf
3R-R'}| +\frac{5\mu}{8\alpha^2}])-2 \sinh(\frac{3\mu}{8}|{\bf
3R-R'}|)\} +(R\leftrightarrow R') $ \\ &\\ \hline &\\ $C$ &
$\frac{\exp(\frac{\mu^2}{4\alpha^2})} {\frac{3}{2}|{\bf
R+R'}|}\{\exp(-\frac{3 \mu}{4}|{\bf R+R'}|)
erf(\alpha[\frac{3}{4}|{\bf R+R'}|-\frac{\mu}{2\alpha^2}])
+\exp(\frac{3 \mu}{4}|{\bf R+R'}|) $ \\ $ $&
$erf(\alpha[\frac{3}{4}|{\bf R+R'}|+\frac{\mu}{2\alpha^2}]) - 2
\sinh(\frac{3\mu}{4}|{\bf R+R'}|)\}+(R\leftrightarrow R') $ \\ &\\
\hline &\\ $C^{'}$ & $\alpha\sqrt{\frac{2}{\pi}}-\mu\cdot 
erfc(\frac{\mu} {\sqrt{2}\alpha}) \exp[\frac{\mu^2}{2\alpha^2}]$
\\
&\\ \hline \hline
\end{tabular}
\end{center}
\caption{The expectation of
$\langle\frac{e^{-\mu r_{lm}}}{r_{lm}}P^0_{ij}
\rangle/K^{0}({\bf R},{\bf R'})$.
$K^{0}({\bf R},{\bf R'})$=$(\frac{3^{9}
\alpha^{6}}{\pi^{3}2^{12}})^{1/2}
\exp[-\frac{15}{16}\alpha^{2}(R^{2}+R'^{2})+\frac{8}{9}\alpha^{2}{\bf R}\cdot
{\bf R'}]$. To get $\langle \frac{1}{r_{ij}}P^{0}_{ij}\rangle$ simply take $\mu$=0.}
\end{table}
\vspace{2.0cm}

\newpage
\begin{figure}[tbp]
\begin{center}
\epsfig{file=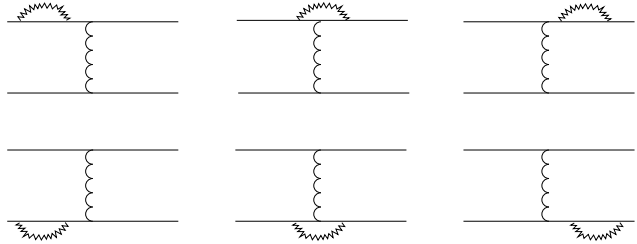,width=3.3in}
\end{center}
\caption{Penguin-like diagrams for $O(\alpha\alpha_s)$ isospin
violation arising from QED-QCD interference. The sharply varying,
smoothly varying, and  solid lines  refer to
photons, gluons, and quarks, respectively.}
\label{fig1}
\end{figure}
\begin{figure}[tbp]
\begin{center}
\epsfig{file=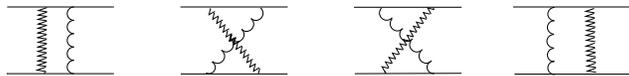,width=3.3in}
\end{center}
\caption{Corresponding box-like graphs. Notations same as Fig. 1.}
\label{fig2}
\end{figure}

\begin{table}
\begin{center}
\begin{tabular}[thb]{||c|c|c|c|c|c||c|c|c||c||}\hline
\hline &&&&&&&&&\\ &$\delta m_q$ & box (el) & box (mag) & pen (el)
& pen (mag) & subtotal & EC & CDEOPE &total \\ &&&&&&&&&\\ \hline
&&&&&&&&&\\ $\delta a_{CSB}$ & -1.83 & -0.05& -1.09& 0.46& 3.48 &
0.97& 0.38 & 0 & 1.35 \\ &&&&&&&&&\\
 \hline
&&&&&&&&&\\ $\delta a_{CIB}$ & 0 & -1.63& -1.19 & 0& 0 & -2.82 &
0.38 & 0.08 & -2.36 \\ &&&&&&&&&\\
 \hline
 \hline
\end{tabular}
\caption{The quark contributions (in fm) to $\delta a_{CSB} =
a_{pp} - a_{nn}$, $\delta a_{CIB} =
\frac{1}{2}(a_{pp}+a_{nn})-a_{np}$. The notation in the entries
is, e.g., pen (el) refers to the electric effect induced by the
penguin-like graphs. EC and CDEOPE refer to the effect of
quark-exchange  Coulomb and OPE interaction. Here $\Delta \Sigma^*
=-4.40MeV$, $\delta m_q=-2.97MeV$,
 $A=-1.96$, $B=7.90$, $C=9.41$, $D=-13.28$.}

\end{center}

\end{table}

\begin{table}
\begin{center}
\begin{tabular}[thb]{||c|c|c|c|c|c||c|c|c||c||}\hline
\hline &&&&&&&&&\\ &$\delta m_q$ & box (el) & box (mag) & pen (el)
& pen (mag) & subtotal & EC & CDEOPE &total \\ &&&&&&&&&\\ \hline
&&&&&&&&&\\ $\delta a_{CSB}$ & -0.52 & -0.03& -1.00& 0.32& 2.28 &
1.05& 0.38 & 0 & 1.43 \\
&&&&&&&&&\\
 \hline
&&&&&&&&&\\ $\delta a_{CIB}$ & 0 & -1.48& -1.09 & 0& 0 & -2.57 &
0.38 & 0.08 & -2.11 \\
&&&&&&&&&\\
 \hline
 \hline
\end{tabular}
\caption{The notation is same with Table 2. Here $\Delta \Sigma^*
=-4.75MeV$, $\delta m_q=-0.85MeV$,
 $A=-1.78$, $B=7.20$, $C=6.91$, $D=-8.71$.}

\end{center}

\end{table}
\begin{table}
\begin{center}
\begin{tabular}[thb]{||c|c|c|c|c|c||c|c|c||c||}\hline
\hline &&&&&&&&&\\ &$\delta m_q$ & box (el) & box (mag) & pen(el)
& pen(mag) & subtotal & EC & CDEOPE &total \\ &&&&&&&&&\\ \hline
&&&&&&&&&\\ $\delta a_{CSB}$ & 0.76 & -0.03& -0.91& 0.22& 1.10 &
1.14& 0.38 & 0 & 1.52 \\ &&&&&&&&&\\
 \hline
&&&&&&&&&\\ $\delta a_{CIB}$ & 0 & -1.33& -0.98 & 0& 0 & -2.31 &
0.38 & 0.08 & -1.85\\
&&&&&&&&&\\
 \hline
 \hline
\end{tabular}
\caption{The notation is same with Table 2. Here $\Delta \Sigma^*
=-5.10MeV$, $\delta m_q=1.27MeV$,
 $A=-1.61$, $B=6.51$, $C=4.41$, $D=-4.14$.}


\end{center}

\end{table}
\begin{table}
\begin{center}
\begin{tabular}[thb]{||c|c|c|c|c|c||c|c|c||c||}\hline
\hline &&&&&&&&&\\ &$\delta m_q$ & box (el) & box (mag) & pen (el)
& pen (mag) & subtotal & EC & CDEOPE &total \\ &&&&&&&&&\\ \hline
&&&&&&&&&\\ $\delta a_{CSB}$ & 1.98& -0.02& -0.81& 0.10& -0.10 &
1.15& 0.38 & 0 & 1.53 \\
&&&&&&&&&\\
 \hline
&&&&&&&&&\\ $\delta a_{CIB}$ & 0 & -1.19& -0.87 & 0& 0 & -2.06 &
0.38 & 0.08 & -1.60\\
&&&&&&&&&\\
 \hline
 \hline
\end{tabular}
\caption{The notation is same with Table 2. Here $\Delta \Sigma^*
=-5.45MeV$, $\delta m_q=3.39MeV$, $A=-1.43$, $B=5.81$, $C=1.90$,
$D=0.43$.}
\end{center}

\end{table}
\begin{table}
\begin{center}
\begin{tabular}[thb]{||c|c|c|c|c|c||c|c|c||c||}\hline
\hline &&&&&&&&&\\ &$\delta m_q$ & box (el) & box (mag) & pen (el)
& pen (mag) & subtotal & EC & CDEOPE &total \\ &&&&&&&&&\\ \hline
&&&&&&&&&\\ $\delta a_{CSB}$ & 3.24& -0.03& -0.71& -0.02& -1.30 &
1.18& 0.38 & 0 & 1.56 \\ &&&&&&&&&\\
 \hline
&&&&&&&&&\\ $\delta a_{CIB}$ & 0 & -1.05& -0.77 & 0& 0 & -1.82 &
0.38 & 0.08 & -1.36\\
&&&&&&&&&\\
 \hline
 \hline
\end{tabular}
\caption{The notation is same with Table 2. Here $\Delta \Sigma^*
=-5.80MeV$, $\delta m_q=5.51MeV$, $A=-1.26$, $B=5.11$, $C=-0.60$,
$D=4.99$}
\end{center}

\end{table}
\end{document}